\documentclass[prl,twocolumn,superscriptaddress]{revtex4}

\usepackage{graphicx}
\usepackage[normalem]{ulem} %need for strikethrough
\usepackage{color} %needed for coloring of \todo etc.

\begin{document}

\title{Shot-noise limited monitoring and phase locking\\ of the motion of a single trapped ion}

\author{P.~Bushev}
\affiliation{Physikalisches Institut, Karlsruher Institut f\"{u}r Technologie, D-76128
Karlsruhe, Germany}
%\affiliation{DFG-Center for Functional Nanostructures (CFN), D-76128 Karlsruhe, Germany}

\author{G.~H\'{e}tet}
\affiliation{Institute for Experimental Physics, University of Innsbruck, A-6020
Innsbruck, Austria} \affiliation{Institute for Quantum Optics and Quantum Information of
the Austrian Academy of Sciences, A-6020 Innsbruck, Austria}

\author{L.~Slodi\v{c}ka}
\affiliation{Institute for Experimental Physics, University of Innsbruck, A-6020
Innsbruck, Austria}

\author{D.~Rotter}
\affiliation{Institute for Experimental Physics, University of Innsbruck, A-6020
Innsbruck, Austria}

\author{M.~A.~Wilson}
\affiliation{Institute for Experimental Physics, University of Innsbruck, A-6020
Innsbruck, Austria}

\author{F.~Schmidt-Kaler}
\affiliation{QUANTUM, Institut f\"{u}r Physik, Universit\"{a}t Mainz, D-55128 Mainz, Germany}

\author{J.~Eschner}
\affiliation{Experimentalphysik, Universit\"{a}t des Saarlandes, D-66123 Saarbr\"{u}cken,
Germany}

\author{R.~Blatt}
\affiliation{Institute for Experimental Physics, University of Innsbruck, A-6020
Innsbruck, Austria} \affiliation{Institute for Quantum Optics and Quantum Information of
the Austrian Academy of Sciences, A-6020 Innsbruck, Austria}

\date{\today}

\begin{abstract}
We perform high-resolution real-time read-out of the motion of a single trapped and
laser-cooled Ba$^+$ ion. By using an interferometric setup we demonstrate shot-noise
limited measurement of thermal oscillations with resolution of 4 times the standard
quantum limit. We apply the real-time monitoring for phase control of the
ion motion through a feedback loop, suppressing the photon recoil-induced phase
diffusion. Due to the spectral narrowing in phase-locked mode, the coherent ion
oscillation is measured with resolution of about 0.3 times the standard quantum limit.
\end{abstract}
\maketitle

%PACS numbers: 42.50.Lc, 42.50.St, 03.65.Ta

The control of a system often relies on gathering information about its evolution in real
time, and using it in a feedback loop to drive it into a desired state~\cite{WienerBook}.
In classical physics, the accuracy of such feedback operations can in theory be
controlled with infinite precision. In quantum physics, however, there is a fundamental
limit to the amount of knowledge one can gain about a system, with important consequences
for its controllability \cite{WisemanBook, Nori2010, White2010,Haroche2011}. If one observes the
motion of a body, the uncertainty in a position measurement inevitably arises due to the
Heisenberg relation. A measurement of its position disturbs its momentum, leading to an
overall blur of the observed motion. The resulting displacement uncertainty is minimized
in the so-called ``standard measurement procedure" which leads to the standard quantum
limit (SQL) \cite{BraginskyBook}. In the case of a harmonic oscillator, the SQL is equal
to the position variance of its ground state $\Delta x_{SQL}=\sqrt{\langle 0|r^2|0
\rangle}=\sqrt{\hbar/2M\omega}$, where $M$ is the mass of the oscillator and $\omega$ its
resonant frequency.

Cooling to the quantum mechanical ground state has been demonstrated for elementary
harmonic oscillators such as trapped ions~\cite{Diedrich1989} and electrons
\cite{Gabrielse1999}, as well as for nanomechanical oscillators~\cite{Martinis2010,
Aspelmeyer2011}. The measurement resolution achieved in experiments with mechanical
oscillators of very different sizes typically settles at the level of 5-10$~\Delta
x_{SQL}$, from the kilogram size oscillator in gravitational wave astronomy in the LIGO
experiment~\cite{Abbott2009}, down to various nanomechanical systems~\cite{Schwab2004,
Aspelmeyer2009, Heidmann2010}. Only recently, resolution below the standard quantum limit
has been demonstrated with nanomechanical oscillators~\cite{Lehnert2009, Kippenberg2010}.
In experiments measuring the trajectory of a single atom in real time, the resolution
achieved so far has remained bounded by a few $\Delta x_{SQL}$~\cite{Kimble2000,
Rempe2000}.
\begin{figure}[ht!]
\includegraphics[width=1\columnwidth]{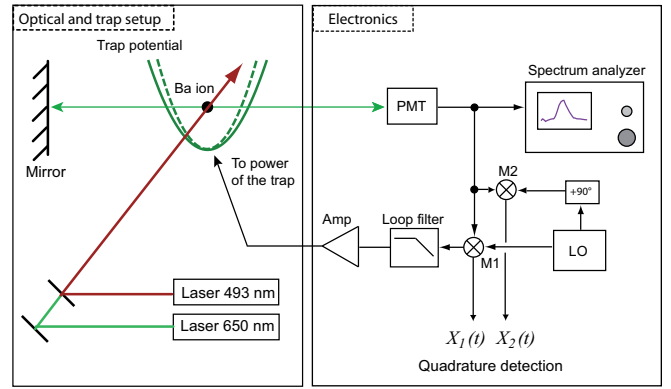}
\caption{(Color online)~Optical and electronic setup for continuous read-out of ion
oscillation and feedback operation. LO denotes local oscillator, PMT stands for
photomultiplier (in photon counting mode), M1 and M2 are radio frequency mixers, Amp is
an amplifier. The parabola represents the trapping potential confining the
ion.}\label{Setup}
\end{figure}

In this letter we present high-resolution real-time read-out of the motion of a single
trapped and laser-cooled barium ion. The relevant oscillation frequency of one of its
macromotion degrees of freedom is $\omega_0/2\pi\sim$~1 MHz, corresponding to a ground
state wave function with a spatial extension of $\Delta x_{SQL}\sim 6$~nm. Using an
interferometric setup~\cite{Eschner2001, Bushev04} we are able to monitor the trajectory
of the ion motion with $4\Delta x_{SQL}$ resolution, only limited by the resonance
fluorescence noise. We furthermore demonstrate efficient suppression of the photon
recoil-induced phase diffusion of the single-ion motion, by phase locking it to an
external oscillator. The motion of the locked oscillator is detected with resolution
below the SQL.

Figure~\ref{Setup} shows a schematic of the set-up. We use a single $^{138}\rm{Ba}^+$ ion
which is held in a miniature Paul trap and continuously laser-excited and -cooled on its
$S_{1/2}$ to $P_{1/2}$ transition at 493~nm~\cite{Raab2000}. The ion oscillates in three
non-degenerate modes with frequencies of about 1~MHz, 1.2~MHz, and 2~MHz. A
retro-reflecting mirror $L=25$~cm away from the trap and a lens (not shown) are arranged
such that they image the ion onto itself. The 493~nm fluorescence of the ion which is
scattered directly towards the photomultiplier (PMT) interferes with the part of the
fluorescence retro-reflected from the mirror. Scanning the ion-mirror distance then leads
to interference fringes at the position of the PMT with a contrast of up to
73\%~\cite{Eschner2001}.

\begin{figure}[ht!]
\includegraphics[width=1\columnwidth]{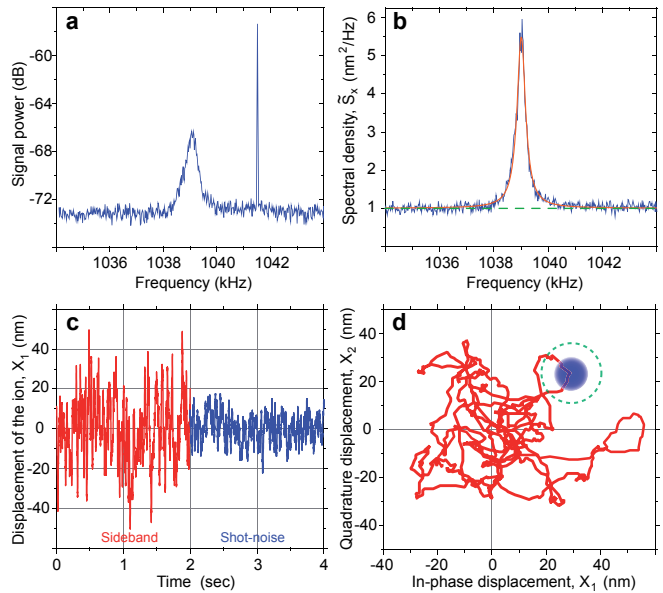}
\caption{(Color online)~Quantum limited detection of the ion's oscillations.
\textbf{(a)}~Spectrum of the photocurrent showing motional sideband (wide resonance line)
and coherent excitation (sharp peak) used for calibration of the ion's motion. The
spectrum was measured with 30 Hz resolution bandwidth (RBW). \textbf{(b)}~The blue curve
is the calibrated spectral density of the ion displacement $\tilde{S}_x(f)$. The red
solid line is a Lorentzian fit, as described by Eq.~(\ref{Sx}). The precision of the
position measurement is set by the shot noise background shown as a dashed green line.
\textbf{(c)}~Time evolution of the in-phase homodyne signal $X_1(t)$ measured when the
local oscillator is tuned in resonance with the ion's oscillation (red curve,
$T=0~\textrm{to}~2$~sec), and when it is tuned off-resonant detecting shot noise only
(blue curve, $T=2~\textrm{to}~4$~sec). The data is smoothed with a 30~Hz low-pass filter,
and the sampling rate was (0.4 ms)$^{-1}$. \textbf{(d)}~Phase plot $(X_1,X_2)(t)$ of the
motion for 300~ms (red line), using data as in (c). The blue circle shows the uncertainty
of any point of the smoothed trajectory due to the shot noise. The green circle shows the
resolution of the unsmoothed trajectory as derived from the spectral measurement in (b).
}\label{Motion}
\end{figure}

For continuous read-out of the ion position, the mirror position is fixed in such a way
that the ion stays at the midpoint of the fringes~\cite{Bushev04}. Then the motion of the
ion leads to a modulation of the intensity of the scattered light, i.e. a modulation in
the arrival times of the photocounts. When the signal from the PMT is measured with an RF
spectrum analyzer, the motion along the trap axis at frequency $\omega_0 = 2\pi \times
1.039$~MHz appears as a clear resonance line above the shot noise background, as shown in
Fig.~\ref{Motion}(a). Using a semiclassical picture for the detected light and
considering the trapped ion as a point particle oscillating with instantaneous
displacement $x(t)$ along the trap axis, the detected photocurrent $i(t)$ is
\begin{equation}\label{photocurrent}
i(t) \propto I_0 \left[ 1 + V \sin(2 k x(t) \cos\Theta) \right] + \delta I(t)~.
\end{equation}
Here $I_0$ is the mean fluorescence rate in counts/sec, $V$ is the visibility of the
interference fringes limited by imperfect optics and imaging as well as by the motion
along other trap axes, $k$ is the wavevector of the 493~nm light, $\Theta = 55^{\circ}$
is the projection angle between trap axis and optical axis, and $\delta I(t)$ represents
the fluorescence shot noise. The single Ba$^+$ ion is trapped in the Lamb-Dicke regime,
therefore Eq.~(\ref{photocurrent}) can be rewritten as
\begin{equation}\label{photocurrent_linear}
i(t) \propto I_0 \left( 1 + 2 V k x(t) \cos\Theta \right) + \delta I(t)~.
\end{equation}
The spectrum of the photocurrent detected by the RF spectrum analyzer consists of three
terms,
\begin{equation}\label{photocurrent_spectrum}
\langle i^2(\omega) \rangle \propto I_0+2\pi I_0^2\delta(\omega) +4I_0^2 V^2 k^2
S_{x}(\omega) \cos^2\Theta~.
\end{equation}
The first term corresponds to the shot-noise which scales linearly with $I_0$, the second
one is the DC-offset or mean intensity level which is outside our observation frequency
range. The last term reveals the power spectral density of the ion motion, $S_x(\omega)$.
It is calculated using the Fluctuation-Dissipation Theorem (FDT)~\cite{CallenWeltonFDT,
Heidmann2000, Schwab2004, Abbott2009} for a harmonic oscillator coupled to a thermal bath
with temperature $T$, which yields
\begin{equation}\label{Sx}
S_{x}(\omega)=\frac{4k_{B}T \gamma}{M} \cdot
\frac{1}{(\omega_{0}^2-\omega^2)^{2}+\gamma^{2}\omega^2}~,
\end{equation}
where $M$ is the mass of the ion and $\gamma$ is the damping rate of the oscillation. In
our case, the measured resonance width $\gamma = 2\pi \times 380$~Hz gives directly the
cooling rate of the trapped ion~\cite{Raab2000}. The area under the resonance curve is
proportional to the thermal energy of the oscillator,
\begin{equation}\label{meanx2}
\langle x^2 \rangle = \frac{1}{2\pi} \int_0^{\infty} S_x(\omega) d\omega =
\int_0^{\infty} \tilde{S}_x(f) df~.
\end{equation}

To calibrate the power spectral density $\tilde{S}_x(f)$ measured on the spectrum
analyzer in terms of displacement, we apply a weak sinusoidal signal to one of the trap
electrodes~\cite{BushevPhD, Bushev06}. This weak excitation is detuned by 2.5~kHz from
$\omega_0$ and coherently excites the ion motion, as shown in Fig.~\ref{Motion}(a). The
relation between the motional amplitude of this coherent oscillation and the spectrum
analyzer signal is found by measuring the change of the (time-averaged) fringe contrast
at different excitation levels. Using Eq.~(\ref{photocurrent}), we then obtain the
calibrated spectral density for the ion motion in nm$^2$/Hz, as shown in
Fig.~\ref{Motion}(b).
The resonance line is well fitted using Eq.~(\ref{Sx}), which yields through
Eq.~(\ref{meanx2}) the mean (r.m.s.) amplitude of the ion excursion $\langle x(t)^2
\rangle^{1/2}=51$~nm. This value is higher than the one estimated from the Doppler
cooling limit, $x_D\approx 27$~nm, set by the linewidth $\Gamma = 2\pi \times 20.4$~MHz
of the $S$ to $P$ cooling transition. That is because the laser detunings and intensities
chosen to maximize the signal-to-noise ratio do not provide optimal cooling conditions.

From figure~\ref{Motion}(b) we determine the achievable resolution of the position
measurement, i.e. the minimum attainable uncertainty of any point of the trajectory when
the ion's motion is monitored in real time. It is set by the shot-noise pedestal with
spectral density $S_{SN}=1.0$~nm$^2$/Hz shown by the dashed line in Fig.~\ref{Motion}(b),
and corresponds to an excursion of the ion motion of 24~nm, or $4 \Delta x
_{SQL}$~\cite{Schwab2004, Abbott2009}. If the oscillation of the ion were aligned with
the optical axis, the resolution would be improved by a factor of $\cos\Theta$ to
2.3$\Delta x _{SQL}$.

To monitor the ion motion, we employ the homodyne technique depicted in Fig.~\ref{Setup}
(see also~\cite{Briant2003}). Part of the detected PMT signal is mixed with a local
oscillator (LO) on two RF-mixers M1 and M2 with $\pi/2$ phase shift between them, and the
two outputs are low-pass filtered. The resulting signals $X_1(t)$ and $X_2(t)$ are given
by
\begin{equation}\label{movavg}
X_i(t)=\int_{-\infty}^t H(t-\eta)x_i(\eta)d \eta~,~(i=1,2),
\end{equation}
where $x_1(t)$ and $x_2(t)$ are the in-phase and quadrature components of the ion motion
(including shot noise), and $H(t-\eta)$ is the time response function of the filter.

A time trace of the $X_1$-component, with the LO set on- and off-resonant with the
oscillation, is shown in Fig.~\ref{Motion}(c). Here a 4-pole filter with $\sim$30~Hz
cut-off frequency is used that transmits about 10\% of the total energy of the sideband;
the smoothed trajectory illustrates the method better. The acquisition time resolution is
set to 0.4 ms. The measured standard deviations are 15~nm on and 7~nm off resonance. In
order to compare them to the oscillator excursion of 51~nm and the resolution of 24~nm as
determined from Fig.~\ref{Motion}(b), they need to be scaled with the square root of
oscillator bandwidth~(380~Hz) over filter bandwidth~(30~Hz), which leads to very good agreement. A phase
plot of the sideband signal, $X_1(t)$ versus $X_2(t)$, is shown during 300 ms in
Fig.~\ref{Motion}(d), exhibiting photon recoil-induced phase diffusion. The circles
indicate the uncertainty on this smoothed trajectory (blue) and the uncertainty for
an optimal filter, i.e. the resolution (green).

We now use the homodyne signal in full real-time conditions, i.e. with a low-pass filter
that matches the oscillator bandwidth. As an application, we seek to drive the ion motion
into a defined coherent state, by phase-locking it to a local oscillator. To this end,
the output of mixer M1 is low-pass filtered with 300~Hz cut-off frequency and amplified,
thus providing an output proportional to the phase lag between the ion's vibration and
the LO. This signal is used to control the intensity of the trap drive in order to change
the trap frequency (see Fig.~\ref{Setup}), thereby counteracting the phase diffusion due
to photon recoils. This operation is reminiscent of a classical phase-locked loop (PLL),
whereby the vibration of the single ion plays the role of the voltage-controlled
oscillator, synchronized to the phase of the LO reference.

\begin{figure}[ht!]
\includegraphics[width=1\columnwidth]{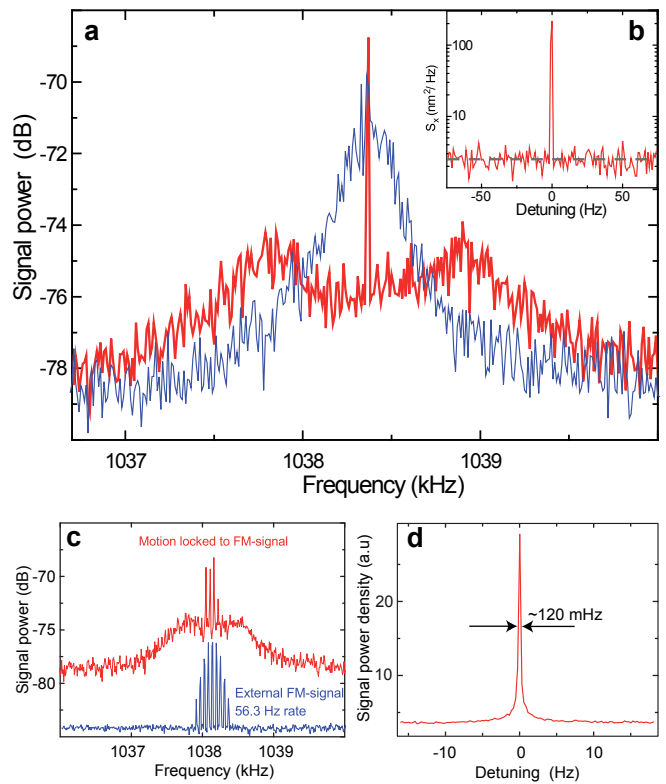}
\caption{(Color online) Phase locking of a trapped ion's vibration. (a)~Motional sideband
spectra when the PLL is off (blue curve) and when the PLL is on (red curve), measured
with RBW = 10~Hz. (b)~Calibrated spectral density $\tilde{S}_x(f)$ around central peak of
phase-locked ion motion, measured with 1~Hz RBW. The dashed line indicates the residual
noise floor of 2.6~nm$^2$/Hz.(c)~Sideband spectrum when the ion is locked to an FM signal
with modulation frequency 56.3~Hz and modulation index 1 (red curve). The spectrum of the
FM reference is also shown (blue curve). (d)~Center peak of phase locked ion motion
measured with 50~mHz resolution, using FFT of the $g^{(2)}$ correlation
function~\cite{Rotter2008}. \label{PLL}}
\end{figure}

To demonstrate the operation of the PLL, the vibrational mode with 1.04~MHz central
frequency is first locked to a sinusoidal LO. We show the relevant spectra with and
without feedback in Fig.~\ref{PLL}(a). Without feedback the sideband spectrum has a
Lorentzian shape with about 500~Hz width, determined by the equilibrium between laser
cooling and recoil heating. In contrast, with the feedback loop closed, phase locking is
clearly observed: the spectrum consists of a narrow central peak and two side-lobes. The
peak contains about 1/8 of the total motional energy, and phase noise is suppressed over
more than 700~Hz bandwidth, indicating that the ion motion follows the LO closely. The
side-lobes indicate the regime where the phase noise is increased, around the oscillation
frequency of the feedback loop (servo bumps). The total energy remains constant, i.e. the
PLL has no damping effect, in contrast to feedback cooling as in \cite{Bushev06}. The
coherent, phase locked ion oscillation is detected with much higher precision than the
free thermal motion due to the narrowing of the resonance line and the detection
bandwidth. As can be seen from Fig.~\ref{PLL}(b), the width of the coherent line is not
yet resolved with 1~Hz resolution bandwidth. Using 1~Hz as the upper limit, we find that
the coherent motion is detected with better than $0.3 \Delta x_{SQL}$ resolution against
the residual phase noise pedestal in closed-loop operation: the spectral density of the
residual noise of 2.6~nm$^2$/Hz corresponds to motion with 1.6~nm mean excursion.

We demonstrate further the potential of the phase lock by the response to a
frequency-modulated (FM) reference, whose spectrum consists of several bands separated by
the modulation frequency. Modulation frequency and index are chosen such that the FM
harmonics do not coincide with 50~Hz noise, and most of the spectral energy falls into
the bandwidth of the loop filter. The resulting spectrum of the motional sideband is
displayed in Fig.~\ref{PLL}(c), with the spectrum of the LO as reference. The spectrum of
ion motion clearly reproduces the three main bands of the FM reference. This measurement
shows, in conjunction with Fig.~\ref{PLL}(b), that we can encode and decode
phase/frequency information into the motion of the single trapped atom with sensitivity
below the standard quantum limit.

To resolve the coherent peak in the locked state even better, we use FFT analysis of the
second order correlation function $g^{(2)}$, following~\cite{Rotter2008}.
Fig.~\ref{PLL}(d) shows the central portion of the FFT spectrum, now using 50~mHz
resolution. The resulting linewidth of the locked signal is 120~mHz. This residual width
is determined mainly by the fact that the time base used in recording $g^{(2)}$ is not
synchronized to the 10~MHz clock used for generating the local oscillator. In principle,
the narrow linewidth would allow deriving a new limit for the resolution of the ion motion,
but calibration proves less reliability for the FFT spectrum than for the one recorded with the rf spectrum analyzer.

In conclusion, we have demonstrated shot-noise limited continuous measurement of the
thermal oscillation of a single ion with 4$\Delta x_{SQL}$ resolution. The resolution
could be moved closer to the standard quantum limit by collecting more fluorescence
light, or by introducing a steeper gradient in the interferometric detection of the
motion. The first method would employ using a lens with higher numerical
aperture~\cite{Sondermann2007, Streed2011}, the second one may be implemented by setting
up the trap inside an optical cavity~\cite{Lange2001, Mundt2002}. We have also realized
phase control of the ion motion, suppressing the photon recoil-induced
phase diffusion through a PLL-type feedback loop. Due to the narrowing of the resonance
line in PLL mode, the coherent ion oscillation is measured with 0.3$\Delta x_{SQL}$
resolution.

The measured trajectory of the ion has three noise contributions arising from its ground state
motion, the shot-noise of the detected fluorescence, and the recoil back-action of the
scattered photons~\cite{Schiller1999}. The latter also determines the steady state of the motion through the laser
cooling process. The PLL control diminishes the shot-noise contribution by narrowing the
observed line. The resulting improvement of the displacement resolution does not affect
the quantum mechanical part of the trajectory $\Delta x$ associated with the zero-point
motion of the ion, and hence does not influence its momentum uncertainty $\Delta p \simeq
\hbar/2\Delta x$. Also the back-action part remains unchanged, since the PLL control stabilizes the phase but does
not alter the temperature of the ion.

The potential of the PLL control has been demonstrated by synchronizing the ion's
oscillation to a modulated reference signal. The phase locking technique may be used to
attain common-mode rejection of recoil effects in photonic atom-atom
interactions~\cite{Cabrillo1999, Eschner2008}. It may also be combined with feedback
cooling~\cite{Bushev06} in order to implement combined amplitude and phase control.

This work has been supported by the Austrian Science Fund (SFB15), the European
Commission (QUEST, HPRNCT-2000-00121; QUBITS, IST-1999-13021), the Spanish MICINN (QOIT,
CSD2006-00019; QNLP, FIS2007-66944), and the "Institut f\"{u}r Quanteninformation GmbH".
P.~B. and J.~E. thank M.~Chwalla, M.~Aspelmeyer and G.~Morigi for stimulating discussions.

\bibliography{PLL_PRL}
\end{document}